%% file: main.tex
\documentclass[pageno]{jpaper}

\usepackage{mathptmx} 
\usepackage{fancyhdr}
\usepackage[normalem]{ulem}
\usepackage[keeplastbox]{flushend}
\usepackage[english]{babel}
\usepackage[utf8]{inputenc}
\usepackage{algorithm}
\usepackage{float}
\usepackage{wrapfig}
\usepackage[noend]{algpseudocode}
\usepackage{textcomp}
\usepackage[dvipsnames]{xcolor}
\usepackage{xspace}
\usepackage{subfig}
\usepackage{booktabs}
\usepackage{enumitem}
\usepackage{cite}
\usepackage{graphicx}
\usepackage{pifont}
\usepackage{setspace}
\usepackage{mwe}
\usepackage{multirow}
\usepackage[compact]{titlesec}
\usepackage{amsmath}
\usepackage{tikz}

\newcommand{\mech}{{SIMDRAM}\xspace} 
\titlespacing\section{0pt}{5pt plus 2pt minus 2pt}{0pt plus 2pt minus 2pt}
\titlespacing\subsection{0pt}{5pt plus 2pt minus 2pt}{0pt plus 2pt minus 2pt}
\titlespacing\subsubsection{0pt}{5pt plus 2pt minus 2pt}{0pt plus 2pt minus 2pt}

\begin{document}

\title{\mech: A Framework for Bit-Serial SIMD Processing Using DRAM\\ \emph{Extended Abstract}}

\newcommand{\affilETH}{$^\diamond$}
\newcommand{\affilCMU}{$^\ddag$}
\newcommand{\affilUIUC}{$^\odot$}
\newcommand{\affilSFU}{$^\star$}

\author{
    \vspace{-15pt}\\
        \thanks{Nastaran Hajinazar and Geraldo F. Oliveira are co-primary authors.}\enspace \scalebox{0.91}{Nastaran Hajinazar}\affilETH\affilSFU~\quad%
        \footnotemark[1]\enspace \scalebox{0.91}{Geraldo F. Oliveira}\affilETH~\quad%
        \scalebox{0.91}{Sven Gregorio}\affilETH~\quad%
        \scalebox{0.91}{João Dinis  Ferreira}\affilETH~\quad%
        \scalebox{0.91}{Nika Mansouri Ghiasi}\affilETH%
    \\%
    \vspace{-12pt}\\%
        \scalebox{0.91}{Minesh Patel}\affilETH~\quad%
        \scalebox{0.91}{Mohammed Alser}\affilETH~\quad%
        \scalebox{0.91}{Saugata Ghose}\affilUIUC~\quad%
        \scalebox{0.91}{Juan Gómez-Luna}\affilETH~\quad%
        \scalebox{0.91}{Onur Mutlu}\affilETH
    \vspace{8pt}\\%
        \it\normalsize \affilETH ETH Z{\"u}rich  \quad  \affilSFU Simon Fraser University \quad \affilUIUC University of Illinois at Urbana--Champaign
    \vspace{-10pt}%
}

\date{}
\maketitle

\thispagestyle{empty}

\input{1_introduction.tex}


{
  \bstctlcite{IEEEexample:BSTcontrol} 
   \let\OLDthebibliography\thebibliography
  \renewcommand\thebibliography[1]{
    \OLDthebibliography{#1}
    \setlength{\parskip}{0pt}
    \setlength{\itemsep}{0pt}
  }
  \bibliographystyle{IEEEtranS}
  \bibliography{refs}
}

\end{document}

%% file: 1_introduction.tex
\section{Motivation \& Limitations of State-of-the-Art}

The increasing prevalence and growing size of data in modern applications
has led to high costs for computation in traditional computer architectures.
Moving large volumes of data between memory devices (e.g., DRAM) and the CPU across 
bandwidth-limited memory channels can consume more than
60\% of the total energy in modern systems\cite{mutlu2019, boroumand2018google}.
To mitigate these costs, researchers have proposed a new
computing paradigm, known as \emph{processing-in-memory} (PIM).
The key idea of PIM is to move computation closer to where the 
data resides, reducing (and in some cases eliminating) the
need to move data between memory and the processor.

There are two main approaches to PIM~\cite{ghoseibm2019, mutlu2020modern}:
(1)~processing-near-memory, where PIM logic is added to the same die as memory or to the logic layer of 3D-stacked memory~\cite{lee2016simultaneous, ahn2015scalable, nai2017graphpim, boroumand2018google, lazypim, top-pim, gao2016hrl, kim2018grim, drumond2017mondrian, santos2017operand, NIM, PEI, gao2017tetris, Kim2016, gu2016leveraging, HBM, HMC2, boroumand2019conda, hsieh2016transparent, cali2020genasm,Sparse_MM_LiM, NDC_Micro_2014, farmahini2015nda,loh2013processing,pattnaik2016scheduling,akin2016data, hsieh2016accelerating,babarinsa2015jafar,lee2015bssync, devaux2019true}; and (2)~processing-using-memory, which makes use of the operational principles of the memory cells themselves to perform computation by enabling interactions between cells~\cite{Chi2016, Shafiee2016, seshadri2017ambit, seshadri2019dram, li2017drisa, seshadri2013rowclone, seshadri2016processing, deng2018dracc, xin2020elp2im, song2018graphr, song2017pipelayer,gao2019computedram, eckert2018neural, aga2017compute,dualitycache}. Since processing-using-memory operates directly in the memory cells, it benefits from the large internal bandwidth and parallelism available inside the memory arrays, which are significantly higher than those for processing-near-memory solutions.

A common approach for processing-using-memory architectures is to make use of bulk bitwise computation. Many widely-used data-intensive applications (e.g., databases, neural networks, graph analytics) heavily rely on a broad set of simple (e.g., AND, OR, XOR) and complex (e.g., equality check, multiplication, addition) bitwise operations. Ambit~\cite{seshadri2017ambit, seshadri2015fast}, an in-DRAM processing-using-memory accelerator, was the first work to propose exploiting DRAM's analog operation to perform bulk bitwise AND, OR, and NOT logic operations. Inspired by Ambit, many prior works have explored DRAM (as well as NVM) designs that are capable of performing in-memory bitwise operations~\cite{angizi2019graphide, angizi2018imce, ali2019memory, pinatubo2016, gao2019computedram, xin2020elp2im}.
However, a major shortcoming prevents these proposals from becoming widely applicable: 
they support only basic operations (e.g., Boolean operations, addition) and fall short on flexibly supporting new and more 
complex operations. Some prior works propose processing-using-DRAM designs that support more complex operations~\cite{li2017drisa, deng2018dracc}. However, such designs (1)~require significant changes to the DRAM subarray, and (2)~support only a limited and specific set of operations, lacking the flexibility to support new operations and cater to the wide variety of applications that can potentially benefit from in-memory computation. \textbf{Our goal} in this paper is to design a framework that aids the adoption of processing-using-DRAM by efficiently implementing complex operations and providing the flexibility to support new desired operations.

\section{The Proposal}

We propose SIMDRAM, an end-to-end processing-using-DRAM framework that provides the programming interface, the ISA and the hardware support for (1)~efficiently computing \emph{complex} operations, and (2)~providing the ability to implement \emph{arbitrary} operations as required, all in an in-DRAM massively-parallel SIMD substrate.
At its core, we build the SIMDRAM framework around a DRAM substrate that enables two previously-proposed techniques:
(1)~vertical data layout in DRAM, and 
(2)~majority-based logic for computation.

\textbf{Vertical Data Layout.} Supporting bit-shift operations is essential for implementing complex computations, such as addition or multiplication. Prior works show that employing a vertical layout~\cite{batcher1982bit,shooman1960parallel, gao2019computedram,ali2019memory,eckert2018neural, dualitycache} for the data in DRAM,  such that all bits of an operand are placed in a single DRAM column (i.e., in a single bitline), eliminates the need for adding extra logic in DRAM to implement shifting~\cite{deng2018dracc, li2017drisa}. Accordingly, \mech supports efficient bit-shift operations by storing operands in a vertical fashion in DRAM. This provides \mech with two key benefits. First, a bit-shift operation can be performed by simply copying a DRAM row into another row (using RowClone~\cite{seshadri2013rowclone}, LISA~\cite{chang2016low} or FIGARO~\cite{wang2020figaro}). For example, SIMDRAM can perform a left-shift-by-one operation by copying the data in DRAM row $j$ to DRAM row $j+1$. (Note that while SIMDRAM supports bit shifting, we can optimize many applications to avoid the need for explicit shift operations, by simply changing the row indices of the SIMDRAM commands that read the shifted data). Second, \mech enables massive parallelism, wherein each DRAM column operates as a SIMD lane by placing the source and destination operands of an operation on top of each other in the same DRAM column.

\textbf{Majority-Based Computation.} Prior works use majority operations to implement basic logical operations~\cite{seshadri2017ambit,gao2019computedram,seshadri2015fast,li2017drisa} (e.g., AND, OR) or addition~\cite{angizi2019graphide,ali2019memory,gaillardon2016programmable,li2017drisa,deng2018dracc,gao2019computedram}. These basic operations are then used as basic building blocks to implement the target in-DRAM computation. \mech extends the use of the majority operation by directly using the logically complete set of majority (MAJ) and NOT operations to implement in-DRAM computation. Doing so enables \mech to achieve higher performance, throughput, and reduced energy consumption compared to using basic logical operations as building blocks for in-DRAM computation. We find that a computation typically requires fewer DRAM commands using MAJ and NOT than using basic logical operations such as AND, OR, and NOT. 

\section{SIMDRAM Framework}

\mech is the first end-to-end framework for processing-using-DRAM. \mech consists of three key steps to enable a desired operation in DRAM: (1)~building an efficient MAJ/NOT-based representation of the desired operation, (2)~mapping the operation input and output operands to DRAM rows and to the required DRAM commands that produce the desired operation, and (3)~executing the operation. These three steps ensure efficient computation of a wide range of arbitrary and complex operation in DRAM. The first two steps give users the flexibility to efficiently implement and compute any desired operation in DRAM. The third step controls the execution flow of the in-DRAM computation, transparently from the user. We briefly describe these steps.

The goal of the first step is to use logic optimization to minimize the number of DRAM row activations, and therefore the compute latency required to perform a specific operation. Accordingly, for a desired computation, the first step is to derive its \emph{optimized} MAJ/NOT-based implementation from its AND/OR/NOT-based implementation.

The second step translates the MAJ/NOT-based implementation into DRAM row activations. This step includes (1)~mapping the operands to the designated rows in DRAM, and (2)~defining the sequence of DRAM row activations that are required to perform the computation. \mech chooses the operand-to-row mapping and the sequence of DRAM row activations to minimize the number of DRAM row activations required for a specific operation. 

The third step is to program the memory controller to issue the sequence of DRAM row activations to the appropriate rows in DRAM to perform the computation of the operation from start to end. To this end, \mech uses a \emph{control unit} in the memory controller that transparently executes the sequence of DRAM row activations for each specific operation.

\section{System Integration}

To incorporate SIMDRAM into a real system, we address 
three integration challenges as part of our work:
(1)~managing memory with both vertical and horizontal layouts
in a system,
(2)~exposing \mech functionality to programmers and compilers, and
(3)~dealing with potential RowHammer-based security exploits~\cite{kim2014flipping, mutlu2019rowhammer,kim2020revisiting, frigo2020trrespass,mutlu2017rowhammer}.
As part of the support for system integration, we introduce two components.

First, \mech adds a \emph{transposition unit} in the memory controller that transforms the data layout from the conventional horizontal layout to vertical layout (and vice versa), as required, thereby allowing both layouts to coexist. Using the transposition unit, \mech provides the ability to store only the data that is required for in-DRAM computation in the vertical layout. As a result, \mech maintains the horizontal layout for the rest of the data and allows the CPU to read/write its operands from/to DRAM in a horizontal layout and at full bandwidth. 

Second, \mech extends the ISA to enable the user/compiler to communicate with the \mech control unit. These extensions include instructions for (1)~transposing data and (2)~indicating specific operations to be issued by the control unit during in-DRAM execution. 

\section{Key Results and Contributions}

The end-to-end support enables \mech as a holistic approach that facilitates the adoption of processing-using-DRAM. The \mech framework efficiently supports a wide range of operations of different types. In this work, we demonstrate the functionality of the \mech framework using an example set of operations including  (1)~\emph{N}-input logic operations (e.g., AND/OR/XOR of more than 2 input bits); (2)~relational operations (e.g., equality/inequality check, greater than, maximum, minimum); (3)~arithmetic operations (e.g., addition, subtraction, multiplication, division); (4)~predication (e.g., if-then-else); and (5)~other complex operations such as bitcount and ReLU~\cite{goodfellow2016deep}. The SIMDRAM framework is not limited to these operations, and can enable processing-using-DRAM for other existing and future operations. 

We compare the benefits of \mech to different state-of-the-art computing platforms (CPU, GPU, and the Ambit~\cite{seshadri2017ambit} in-DRAM computing mechanism). We comprehensively evaluate SIMDRAM's reliability, area overhead, throughput, and energy efficiency. Our evaluation shows that \mech provides up to $5.1\times$ higher throughput and $2.5\times$ higher energy efficiency compared to Ambit~\cite{seshadri2017ambit} for 16 different operations, while incurring less than 1\% DRAM area overhead. 

We leverage the \mech framework to accelerate seven application kernels from machine learning, databases, and image processing (VGG-13~\cite{simonyan2014very}, VGG-16~\cite{simonyan2014very}, LeNET~\cite{lecun2015lenet}, kNN~\cite{lee1991handwritten}, TPC-H~\cite{tpch}, BitWeaving~\cite{li2013bitweaving}, Brightness~\cite{gonzales2002digital}). 
\mech provides up to $2.5\times$ speedup for the kernels compared to Ambit~\cite{seshadri2017ambit}. Compared to a CPU and a high-end GPU, \mech is $257\times$ and $31\times$ more energy efficient, while providing $93\times$ and $6\times$ higher throughput, respectively. We also evaluate the reliability of \mech under different degrees of manufacturing process variation, and observe that it guarantees correct operation as the DRAM process technology node scales down to smaller sizes. 

We make the following key contributions:
\begin{itemize}[noitemsep,topsep=0pt,parsep=0pt,partopsep=0pt,labelindent=0pt,itemindent=0pt,leftmargin=*]
\item To our knowledge, this is the first work to propose a framework to enable efficient computation of a flexible and wide range of operations in a massively parallel SIMD substrate built via processing-using-DRAM. 
\item \mech provides a three-step framework to develop efficient and reliable MAJ/NOT-based implementations of a wide range of operations.  We design this framework, and add hardware and ISA support, to (1)~address key system integration challenges and (2)~allow programmers to employ new \mech operations without hardware changes.
\item We provide a detailed reference implementation of \mech, including required changes to the user applications, ISA, and hardware.
\item We evaluate the reliability of \mech under different degrees of process variation and observe that it guarantees correct operation as the DRAM technology scales to smaller node sizes.
\end{itemize}